\documentclass[11pt]{article}
\pdfoutput=1

\usepackage{euscript}
\usepackage{amssymb}
\usepackage{amsfonts}
\usepackage{amsbsy}
\usepackage{epsfig}
\usepackage{amsthm}
\usepackage{amscd}
\usepackage{amstext}
\usepackage{verbatim}
\usepackage{amsmath}
\usepackage{marginnote}

\def\ben{\begin{equation}}
\def\een{\end{equation}}
\def\half{{\textstyle{\frac12}}}

\let\a=\alpha  \let\g=\gamma \let\d=\delta 
   \let\k=\kappa

\let\w=\omega

\newcommand{\ba}{\begin{align}}
\newcommand{\ea}{\end{align}}
\newcommand{\nn}{\nonumber \\}
\newcommand{\bi}{\begin{itemize}}
\newcommand{\ei}{\end{itemize}}

\def\be{\begin{equation}}
\def\ee{\end{equation}}
\def\beq{\begin{equation}}
\def\eeq{\end{equation}}

\def\dalemb#1#2{{\vbox{\hrule height .#2pt
       \hbox{\vrule width.#2pt height#1pt \kern#1pt
               \vrule width.#2pt}
       \hrule height.#2pt}}}

\newcommand{\bea}{\begin{eqnarray}}
\newcommand{\eea}{\end{eqnarray}}

\def\R{{{\Bbb R}}}

\def\ocal{{\mathcal{O}}}

\begin{document}

\begin{center}

{ \Large {\bf Metal-insulator transition in holography
}}

\vspace{1cm}

Aristomenis Donos$^\sharp$ and Sean A. Hartnoll$^\flat$
\vspace{0.7cm}

{\small
{\it $^\sharp$ Blackett Laboratory, 
        Imperial College, \\ London, SW7 2AZ, U.K. \\
        \vspace{0.3cm}
       $^\flat$  Department of Physics, Stanford University, \\
Stanford, CA 94305-4060, USA }}

\vspace{1.6cm}

\end{center}

\begin{abstract}

We exhibit an interaction-driven metal-insulator quantum phase transition in a holographic model.
Use of a helical lattice enables us to break translation invariance while preserving homogeneity.
The metallic phase is characterized by a sharp Drude peak and a d.c.{\,}resistivity that increases with temperature.
In the insulating phase the Drude spectral weight is transferred into a `mid-infrared' peak and to energy scales of order the chemical potential. The d.c.{\,}resistivity now decreases with temperature. In the metallic phase, operators breaking translation invariance are irrelevant at low energy scales. In the insulating phase, translation symmetry breaking effects are present at low energies. We find the near horizon extremal geometry that captures the insulating physics.

\end{abstract}

\pagebreak
\setcounter{page}{1}

\begin{figure}[h]
\begin{center}
\includegraphics[height = 70mm]{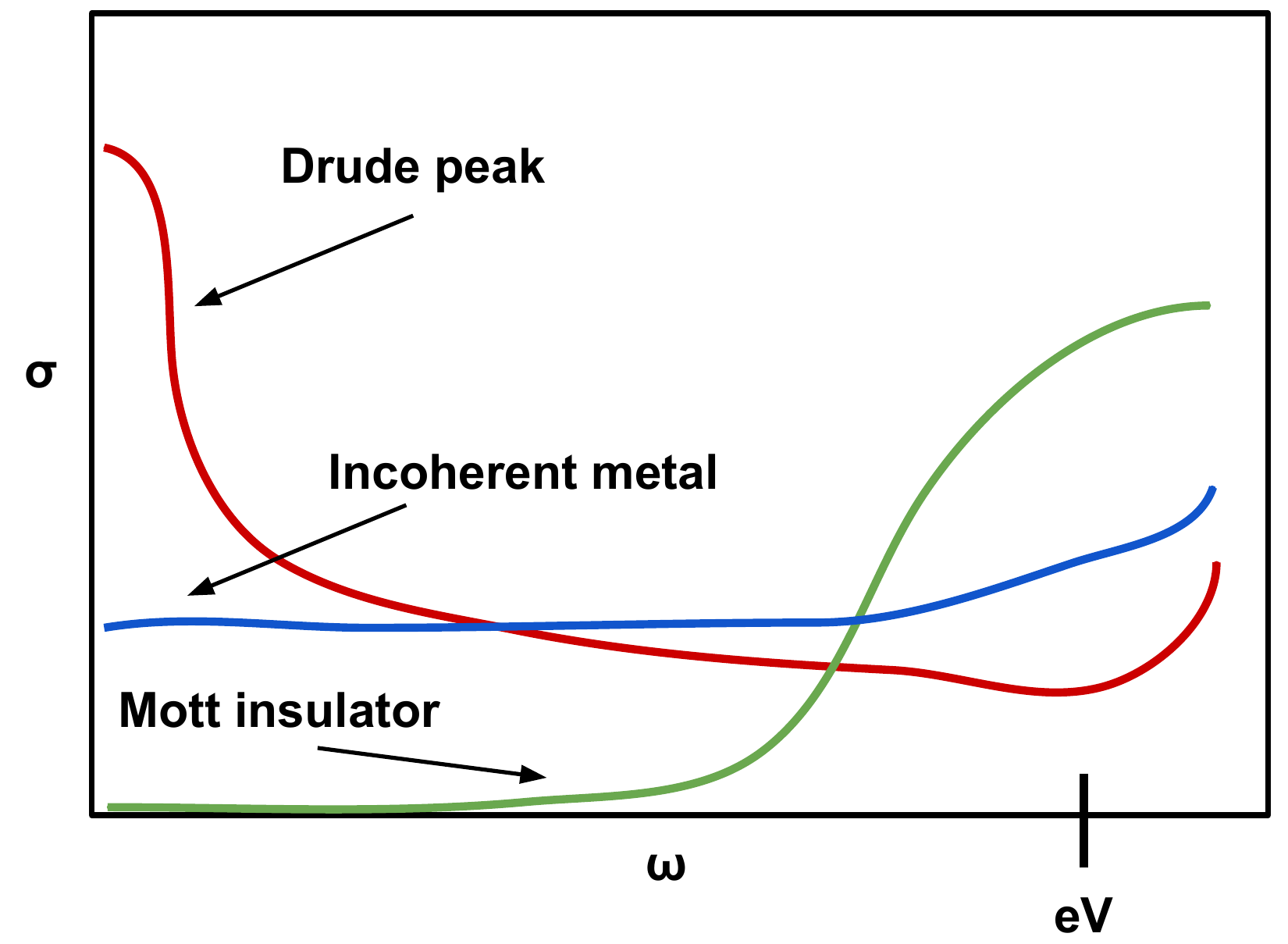}\caption{Progressive loss of Drude coherence in the optical conductivity. From top to bottom on the left: conventional metals, bad metals and Mott insulators.
\label{fig:cartoon}}
\end{center}
\end{figure}

\marginnote{\sc {\small Metal-insulator transitions}}[1cm]  Metal-insulator transitions driven by electron interactions present serious theoretical challenges and yet are a key feature in the phase diagram of several families of materials, including the cuprate superconductors \cite{mitrev,mit2}. As the system transitions to an insulating state, the d.c. conductivity drops dramatically, while in the optical conductivity the Drude peak gives way to a gap in the low energy spectral weight \cite{mitrev, condrev}. In the vicinity of the metal-insulator transition one can encounter `bad metals'. These are characterized by metallic behavior in the absence of a coherent Drude peak and have resistivities exceeding the Mott-Ioffe-Regel saturation limit \cite{hussey, gun, lim}. Figure \ref{fig:cartoon} shows a cartoon of the optical conductivity in conventional metals, bad metals and Mott insulators.  It has been argued that bad metals cannot admit a quasiparticle description because such quasiparticles would necessarily have a mean free path shorter than their Compton wavelength \cite{Emery:1995zz}. In some cases at least, the reorganization of the charge-carrying degrees of freedom, from itinerant to localized excitations, across an interaction-driven metal-insulator transition is likely to involve strong coupling dynamics that is inaccessible to conventional theoretical tools.

\begin{figure}[h]
\begin{center}
\includegraphics[height = 70mm]{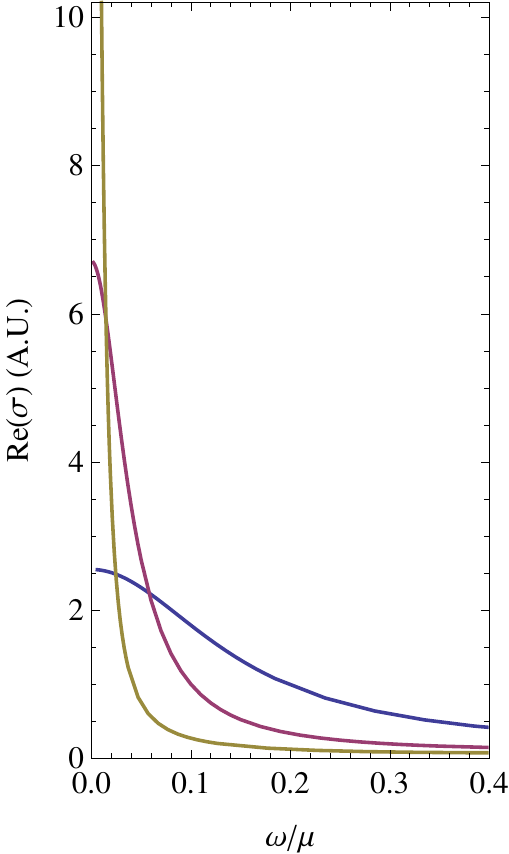}\includegraphics[height = 70mm]{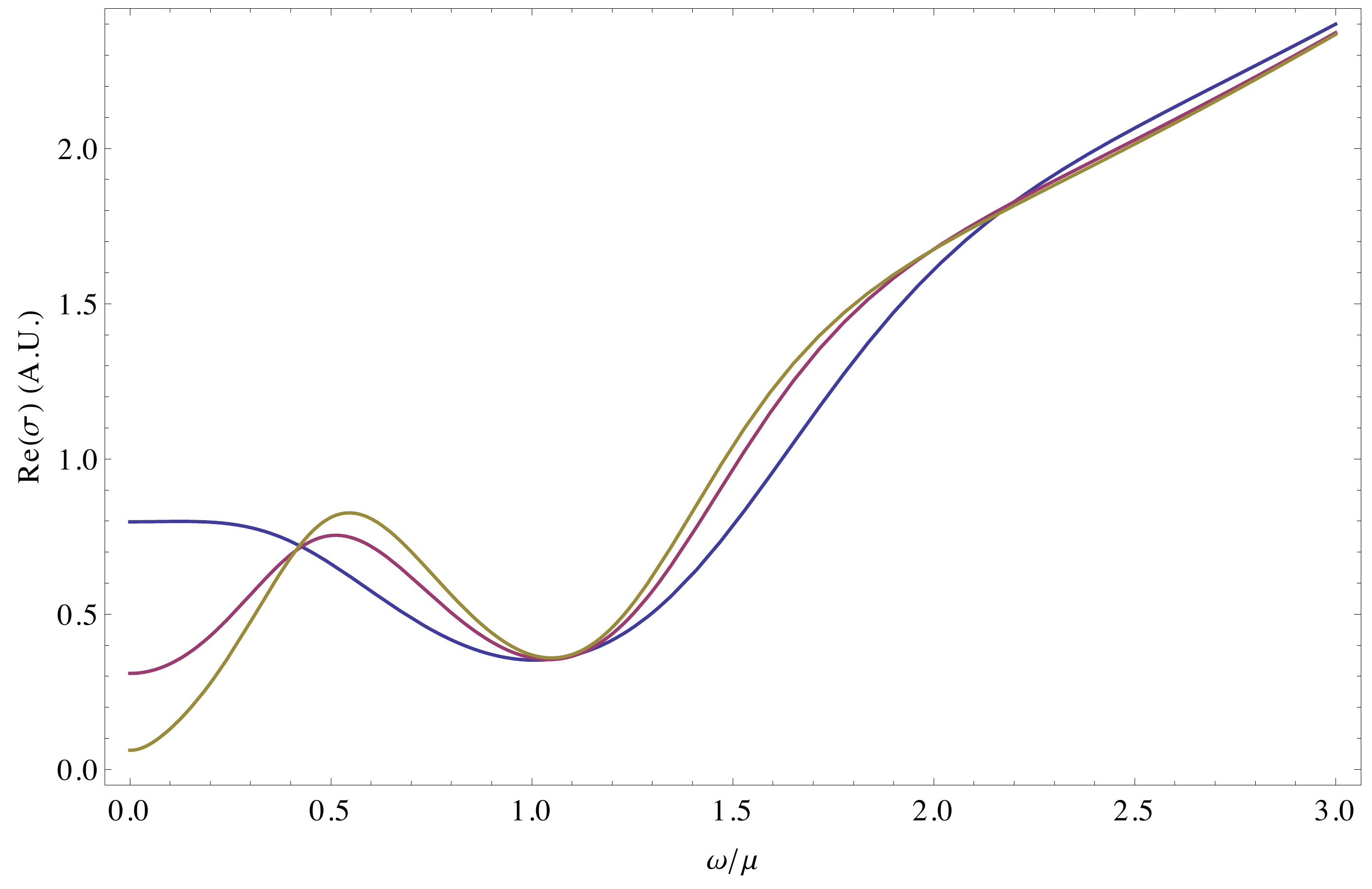}\caption{Optical conductivity in metallic (left) and insulating (right) phases.
The three temperatures shown in each plot are $T/\mu = 0.011, 0.05, 0.12$ (top to bottom in the metallic case, bottom to top in the insulating case). The metallic phase has $p/\mu = 2.5$ and the insulating phase, $p/\mu = 1/\sqrt{2}$. Both have $m^2 = 0$, $\k = 1/\sqrt{2}$ and $\lambda/\mu = 3/2$. The metallic phase exhibits a Drude peak while in the insulating phase low energy spectral weight is transferred into a
`mid-infrared' peak and to interband energy scales. Note the different range of the axes on the two plots.
\label{fig:opt}}
\end{center}
\end{figure}

\begin{figure}[h]
\begin{center}
\includegraphics[height = 70mm]{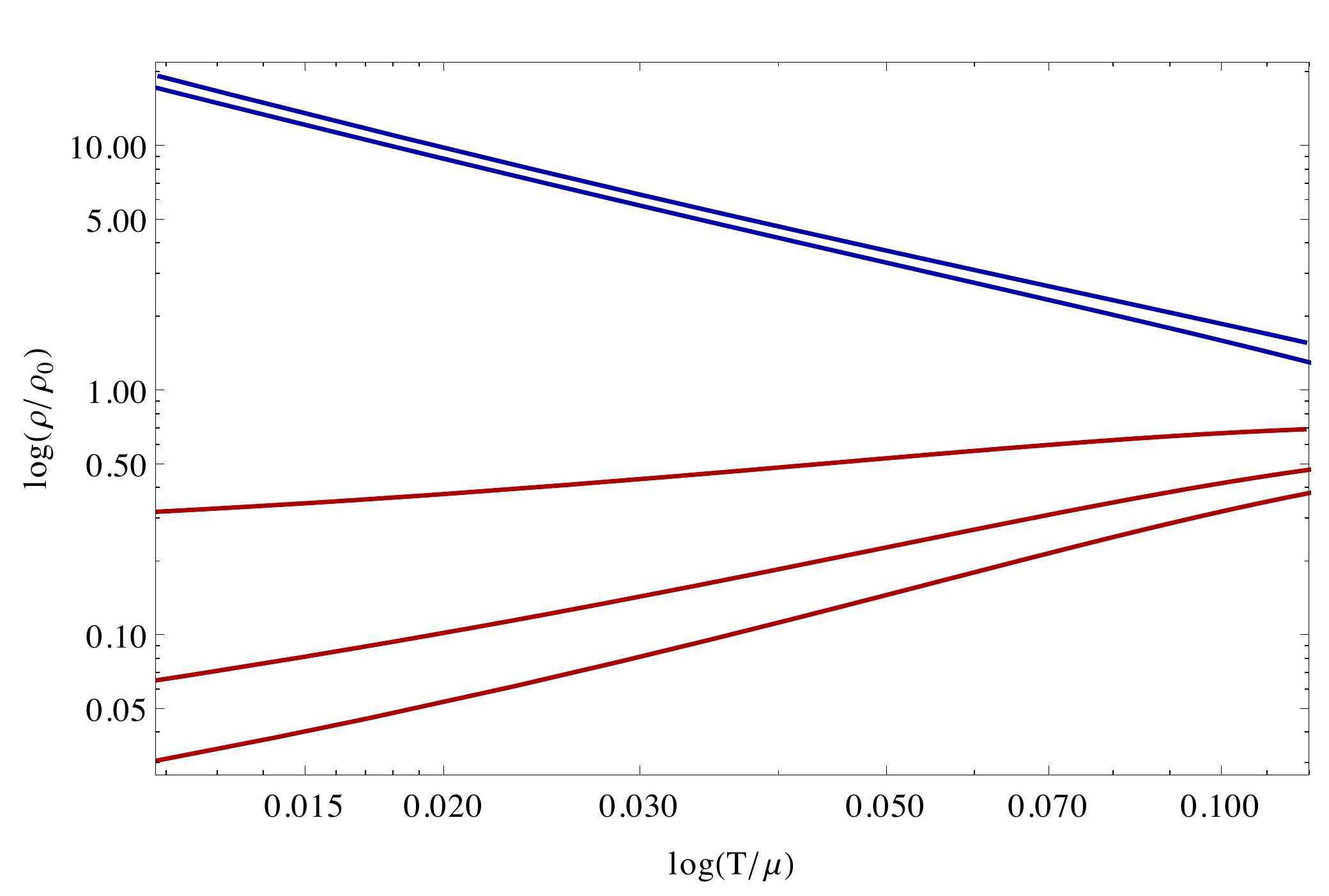}\caption{Log-log plot of the d.c.{\,}resistivity as a function of temperature. From top to bottom, the upper (blue) lines are insulating phases with $p/\mu = 1,1/\sqrt{2}$ and the lower (red) lines are metallic phases with $p/\mu = 2.2, 2.4, 2.5$. As before, $m^2 = 0$, $\k = 1/\sqrt{2}$ and $\lambda/\mu = 3/2$. The resistivity $\rho_0$ is an arbitrary scale.
\label{fig:dc}}
\end{center}
\end{figure}

\marginnote{\sc {\small Holography and transport}}[1cm] The holographic correspondence allows a description of certain strongly interacting media that eschews quasiparticles from the outset \cite{Hartnoll:2011fn}. This approach has been demonstrated to offer new qualitative and quantitative insights into transport in quantum critical particle-hole symmetric theories, e.g. \cite{WitczakKrempa:2012gn}. In this paper we will shall use holographic duality to give an inherently strongly interacting description of a metal-insulator transition in a finite density theory. Our results for the optical (in figure \ref{fig:opt}) and d.c. (in figure \ref{fig:dc}) conductivities capture key qualitative features of the experimental data on metal-insulator transitions, including bad metallic regimes \cite{hussey, gun, lim} and spectral weight transfer \cite{condrev, hussey, orenstein, nicoletti}.

The primary obstacle to the realization of a genuine insulating phase in holography has been the fact that such a phase must break translation invariance. In metallic phases with a sharp Drude peak, the effects of momentum non-conservation are an irrelevant deformation of the continuum theory at low energies that may be treated perturbatively \cite{Hartnoll:2012rj}. Thus the effects of irrelevant disorder \cite{Hartnoll:2007ih, Hartnoll:2008hs} and lattices \cite{Hartnoll:2012rj} on charge transport at low energies and temperatures are easily incorporated into holographic models. The modeling of bad metals and insulators, however, requires the breaking of translation invariance to be treated non-perturbatively at the lowest energy scales. Recent impressive steps in this direction have been taken in \cite{Horowitz:2012ky, Horowitz:2012gs}, where a lattice was fully incorporated into holographic models. However, the regimes studied thus far have all exhibited a conventional Drude peak.

\marginnote{\sc {\small Helical lattices}}[1cm] An important technical innovation of our work will be to break translation invariance while retaining homogeneity of the system.\footnote{David Vegh has independently studied the conductivity in homogeneous systems that are not translation invariant.} This will enable our holographic study of localization physics to operate at the level of ODEs in the bulk, while remaining nonperturbative in the lattice strength, as opposed to the technically involved PDEs that are typically necessary once boundary translation invariance is broken.
Specifically, we turn on the following boundary background fields (couplings)
\be\label{eq:sources}
A^{(0)} = \mu \, dt \,, \qquad B^{(0)} = \lambda \, \w_2 \,.
\ee
Here $\mu$ is the chemical potential for the electric charge while $B^{(0)}$ is a source for a vectorial operator that describes the lattice. The strength of the lattice is given by $\lambda$ while $\w_2$ is one of the three Bianchi VII$_0$ invariant one-forms
\be
\w_1 = dx_1 \,, \qquad \w_2 + i \w_3 = e^{i p x_1} (dx_2  + i dx_3) \,.
\ee
We see that $p$ is the pitch of the helical structure of $B^{(0)}$. The helix rotates in the $x_2$-$x_3$ plane as a function of the $x_1$ direction. This helical lattice breaks translation invariance in the $x_1$ direction while remaining invariant under the non-abelian Bianchi VII$_0$ symmetry algebra. Our discussions of metallic and insulating behavior in the following will always refer to currents in the $x_1$ direction. Currents in the $x_2$ and $x_3$ directions are not relaxed. This extreme anisotropy is a consequence of the symmetries of our microscopic lattice. We adopt this lattice for technical convenience, but it may also be of intrinsic interest in relation to smectic metals \cite{smec}. An introduction to the geometry of the Bianchi VII$_0$ algebra with application to chiral nematics may be found in \cite{Gibbons:2011im}. Such helices have previously been found to emerge spontaneously in holographic studies including \cite{Nakamura:2009tf, Donos:2011ff, Donos:2012wi}. Here we are explicitly sourcing the helical structure as a simplified way to break translation invariance. An RG flow generated by a helical current source was previously considered in \cite{Iizuka:2012iv}.

\marginnote{\sc {\small The bulk model}}[1cm] To extend the sources (\ref{eq:sources}) into the holographic radial bulk dimension $r$ it is sufficient to consider the bulk ansatz
\bea
& A =  a(r) dt \,, \qquad B = w(r) \w_2  \,, \nonumber \\ 
& \displaystyle ds^2  =  - U(r) dt^2 + \frac{dr^2}{U(r)} + e^{2 v_1(r)} \w_1^2 + e^{2 v_2(r)} \w_2^2+ e^{2 v_3(r)} \w_3^2 \,. \label{eq:metric}
\eea
To determine the six radial functions $a,w,v_i,U$ we need an action for the bulk theory. We consider the following simple Einstein-Maxwell-Proca action
\be\label{eq:action}
S = \int d^5x \sqrt{-g}\left(R + 12 - \frac{1}{4} F_{ab} F^{ab} - \frac{1}{4} W_{ab} W^{ab} - \frac{m^2}{2} B_a B^a \right) - \frac{\k}{2} \int B \wedge F \wedge W \,.
\ee
Here $F = dA$ and $W = dB$ are the field strengths. The final Chern-Simons term is not logically essential for our purposes, but will turn out to facilitate finding a metal-insulator transition. We have set the Newton constant, AdS radius and Maxwell couplings to unity. In general all these constants can be scaled into the fields. The wedge product is normalized so that the Chern-Simons term on the ansatz (\ref{eq:metric}) is $\half p \k w^2 a'$. The resulting background equations of motion for $\{ a,w,v_i,U \}$ are listed in the supplementary material.

Near the asymptotic UV boundary at $r \to \infty$ we require that the spacetime tend to $AdS_5$ together with the sources (\ref{eq:sources}). That is, to leading order at large $r$
\be\label{eq:uv}
U = r^2 \,, \qquad v_i = \log r \,, \qquad a = \mu \,, \qquad w = \lambda \,. 
\ee
More important however is the far interior of the spacetime, which will describe the universal low energy physics \cite{Hartnoll:2011fn}. \marginnote{\sc {\small Metallic geometry}}[1cm] Underlying the metal-insulator transition that we will find is the fact that there are different possible far interior geometries at zero temperature. These correspond to different solutions to the Einstein-Maxwell-Proca action. The metallic state at zero temperature will be described by the well known $AdS_2 \times \R^3$ solution to the equations of motion as $r \to 0$, together with specific deformations:
\be\label{eq:ads2}
U = 12 \, r^2 (1 + u_1 r^\d) \,, \qquad v_i = v_o(1 + v_{i\,1} r^\d) \,, \qquad a = 2 \sqrt{6} \, r (1 + a_1 r^\d) \,, \qquad w = w_1 r^\d \,.
\ee
The exponents are determined by the equations of motion and come in pairs $\d_\pm$ with $\d_+ + \d_- = -1$. Each larger value $\d_+$ gives the scaling dimension of a coupling describing a deformation of this semi locally critical IR. We find two marginal operators with $\d_+ = 0$. These correspond to rescaling $x_1$ and $\{x_2,x_3\}$, respectively.
The three remaining operators are found to have
\be\label{eq:delta}
\delta_+ = \left\{1 \,, \quad -\half + \sqrt{\textstyle \frac{1}{12}(3+ m^2) + \frac{1}{12} p^2 e^{- 2 v_o} - \frac{\k}{\sqrt{6}} p \, e^{- v_o} } \,,
\quad -\half + \sqrt{\textstyle \frac{1}{4} + \frac{1}{3} p^2 e^{-2 v_o}} \right\} \,.
\ee
If
\be\label{eq:irrelevant}
\left(2 \sqrt{6}\, \k - p \, e^{-v_o} \right) p \, e^{- v_o} < m^2 \,,
\ee
we can see that all three of these modes are irrelevant, with $\d_+ > 0$. These marginal and irrelevant  modes must be turned on and their coefficients $\{u_1, v_{i\,1}, a_1, w_1\}$ tuned in order to match onto the near boundary behavior (\ref{eq:uv}) upon integrating the equations of motion out to large $r$. In cases where (\ref{eq:irrelevant}) is satisfied, the interior solution is translation invariant, because $w=0$ as $r \to 0$ and the deformation is turned off, and therefore the helical structure is irrelevant. The IR irrelevance of translation invariance breaking operators implies that phases with these interior geometries will be metals with a coherent Drude peak \cite{Hartnoll:2012rj}.

There is no guarantee, however, that the translation invariant solution with irrelevant deformations (\ref{eq:ads2}) is indeed the zero temperature IR geometry. The simplest way in which that geometry can fail to be the true IR is if the mass of the Proca field $B$ and the Chern-Simons coupling are such that the inequality (\ref{eq:irrelevant}) is violated for some $v_o$, at fixed pitch $p$.\footnote{We should require that the asymptotically $AdS_5$ Breitenlohner-Freedman bound is satisfied: $-1 \leq m^2$.} This violation leads to $\d_+ < 0$ for one of the operators about the IR fixed point. The IR datum $v_o$ can be tuned by varying UV data such as the chemical potential. Crossing the bound (\ref{eq:irrelevant}) will trigger a quantum phase transition as the operator becomes relevant and drives an RG flow away from the $AdS_2 \times \R^3$ fixed point. This transition could be first order or continuous, depending on whether the true IR fixed point is close or not to the RG-unstable $AdS_2 \times \R^3$ fixed point when the operator first becomes relevant. This scenario is illustrated in figure \ref{fig:rgflows} below.

If the dimension $\d_+$ becomes complex in (\ref{eq:delta}), this indicates that the interior geometry has become dynamically unstable. These are the inhomogeneous instabilities investigated in e.g. \cite{Nakamura:2009tf, Donos:2012wi}. In this work we are interested in the effects of an explicit helical lattice rather than spontaneously generated helices. We will therefore make sure to stay at values of $m^2,\k$ and $p$ such that $\d_+$ about the metallic $AdS_2 \times \R^3$ background is never complex for any $v_o$. For the massless case that we will focus on below, the condition for stability is seen to be from (\ref{eq:delta}) that $|\kappa| \leq 1/\sqrt{2}$.

\newpage

\begin{figure}[h]
\begin{center}
\includegraphics[height = 70mm]{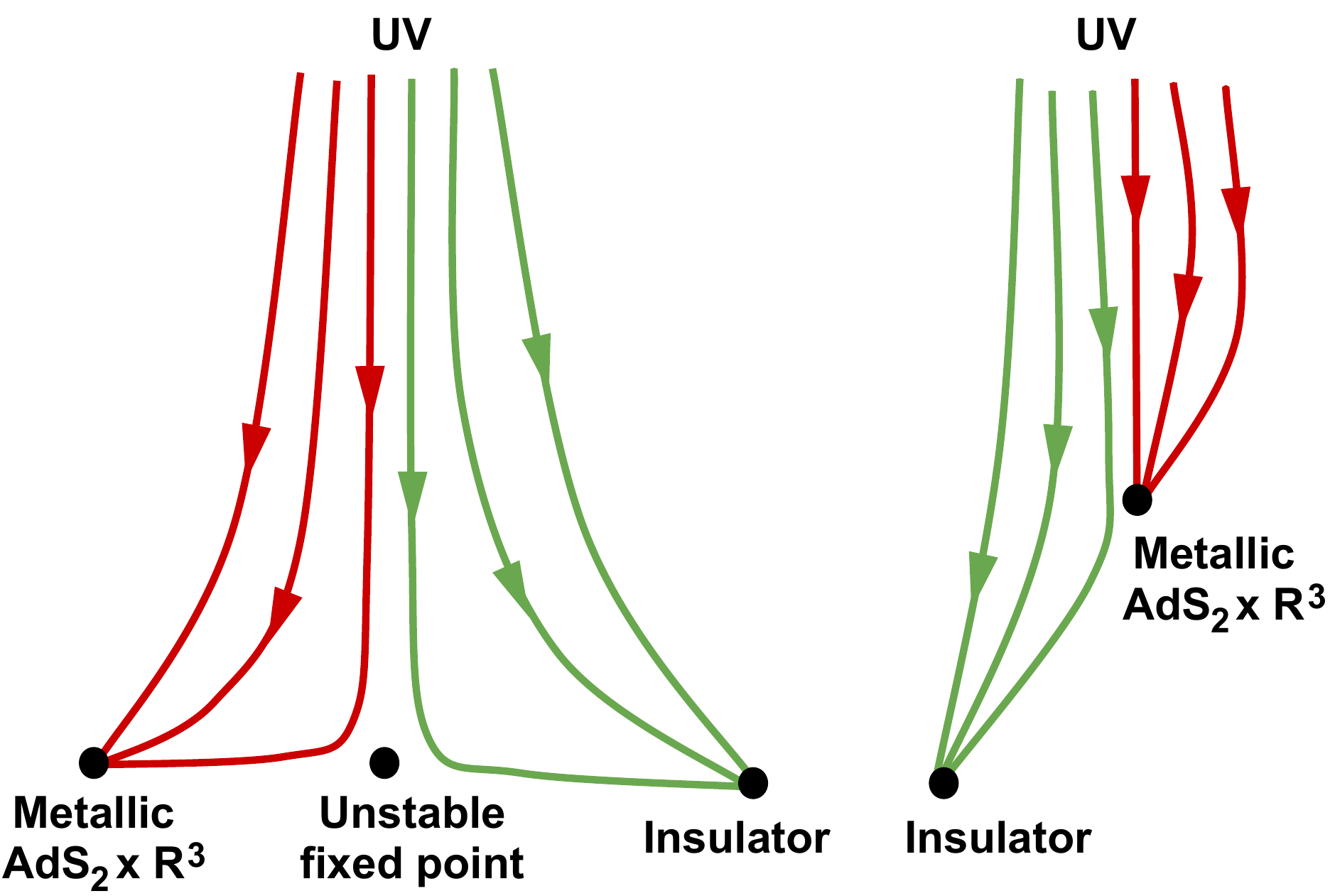}\caption{Two renormalization group flow scenarios that arise in our theories, mediating quantum phase transitions between metallic and insulating phases. In the left plot, the phase transition is mediated by an unstable fixed point, which has relevant operators (that may in addition have complex scaling dimensions). In the right plot, the phase transition occurs when the metallic fixed point itself develops a relevant deformation. One of these two possibilities must occur for the metallic phase to become unstable.
\label{fig:rgflows}}
\end{center}
\end{figure}

\marginnote{\sc {\small Bifuraction geometry}}[1cm]A second way that the IR can be distinct from the translation invariant solution (\ref{eq:ads2}) is if there is a bifurcation in the RG flow as a function of the UV parameters of the system such as the lattice strength. This is the scenario that was found to mediate holographic fractionalization quantum phase transitions in \cite{Hartnoll:2011pp}. The bifurcation fixed point should have a relevant deformation and thus be unstable under RG flow.
If the fixed point mediating the bifurcation is furthermore dynamically unstable (that is, with deformations that have complex scaling dimension), then the metal-insulator quantum phase transition will be first order, otherwise it will be continuous \cite{Hartnoll:2011pp}. The required unstable fixed points are similar to those constructed in \cite{Iizuka:2012iv}; we are able to charge up those solutions using the Chern-Simons term. The Chern-Simons coupling allows the `magnetic' $B$ field to generate an electric $A$ field with a minimal back reaction on the solution. It is simple to see that our theory admits the scaling solution
\be\label{eq:unstable}
U = u_o r^2 \,, \quad v_1 = v_{1o} \,, \quad e^{v_2} = e^{v_{2o}} r^\a \,, \quad e^{v_3} = e^{v_{3o}} r^\a \,, \quad a = a_o r \,, \quad w = w_o r^{\a} \,.
\ee
Plugging this ansatz into the equations of motion, the exponent $\a$ and the various prefactors can be determined numerically by solving algebraic equations, and depend upon the pitch $p$, mass squared $m^2$ and Chern-Simons coupling $\k$. For the case $m = 0$ that we will focus on in the numerics below, we find that for all $p$, the bifurcating solution exists for $|\k| \lesssim 0.57$.

To determine the stability of this fixed point we need to perturb the functions $\{U,v_i,a,w\}$ in (\ref{eq:unstable}) by a power law $r^\d$, analogously to what we did in (\ref{eq:ads2}) above. The operators captured by our bulk theory are found to now correspond to two irrelevant operators with $\d_+ > 0$, one marginal operator with $\d_+ = 0$ and two relevant operators. This fixed point is therefore unstable under RG flow, as anticipated above. Depending on the regime of parameters, one of the relevant operators can have a complex scaling exponent. In the massless case $m=0$, the operator dimensions are found to be independent of the pitch $p$, with the complex scaling exponent present for $|\k| \lesssim 0.3$. Thus depending on the Chern-Simons coupling, a metal-insulator transition mediated by this fixed point is expected \cite{Hartnoll:2011pp} to be first order (for $0 < |\k| \lesssim 0.3 $) or continuous (for $0.3 \lesssim |\k| \lesssim 0.57$). Let use repeat that the solution we are describing at the moment should not be understood as the generic true IR ground state, but rather as an unstable fixed point that mediates the transition between two different phases. This scenario is also illustrated in figure \ref{fig:rgflows}.

\marginnote{\sc {\small Insulating geometry}}[1cm]For the massless case $m=0$, we have found the stable IR geometry that will pertain in the insulating phase. Unlike the metallic IR (\ref{eq:ads2}) and the unstable critical point (\ref{eq:unstable}), the interior geometry in this case is not a scaling solution on its own, but rather describes the leading order terms in an expansion as $r \to 0$. The background to leading order as $r \to 0$ is found to be
\bea
&  & U = u_o r^2  + \cdots \,, \quad e^{v_2} = e^{v_{1o}} r^{-1/3} + \cdots\,, \quad e^{v_2} = e^{v_{2o}} r^{2/3} + \cdots\,, \quad e^{v_3} = e^{v_{3o}} r^{1/3} + \cdots\,, \nonumber \\
&  & a = a_o r^{5/3} + \cdots \,, \quad w = w_o + w_1 r^{4/3} + \cdots \,. \label{eq:insulating}
\eea
We have quoted the subleading term in $w$, as this is necessary to establish consistently the leading order behavior of $a$. We see that the translation invariance breaking `magnetic' background field $W$ does not vanish in this IR background, unlike in the metallic IR (\ref{eq:ads2}). Furthermore,
while the metallic IR solution (\ref{eq:ads2}) has electric flux, $\int_{\R^3} \star F$ through the horizon, the insulating IR background (\ref{eq:insulating}) and the bifurcating scaling solution (\ref{eq:unstable}) have no such flux. The electric field is generated outside the horizon via the Chern-Simons term. We are placing the Chern-Simons term on the right hand side of the Maxwell equation and thinking of it as a source for charge. Our phase transition might therefore be thought of as a non-translationally invariant version of the fractionalization transition discussed in \cite{D'Hoker:2012ej}. There, a Chern-Simons term together with a magnetic field enabled a phase transition between solutions with the charge behind and outside the horizon. While in the models we study in this paper we therefore have that the metals are fractionalized and the insulators `cohesive' \cite{Hartnoll:2012ux}, this is not essential to our setup. We expect similar metal-insulator transitions to occur without the Chern-Simons term, and therefore without the possibility of cohesive phases. In fact, we have found strong evidence for the presence of such transitions (without the Chern-Simons term) using numerics at finite temperature. However, we have not been able to identify analytically the zero temperature IR insulating behavior in these cases. Finally, we should note that the solution (\ref{eq:insulating}) will also describe the zero temperature IR of the helical black holes discussed in \cite{Donos:2012wi}.

The stability of the insulating IR geometry (\ref{eq:insulating}) under RG flow is established by perturbing the background by power law modes as previously. Because the background is not an exact solution, this is a little more subtle than previous cases. Details may be found in the supplementary material. The upshot of the analysis is that the background has no relevant modes and is therefore a stable IR fixed point. By following the irrelevant and marginal modes upwards along the RG flow we can construct asymptotically $AdS_5$ domain wall solutions. We now have all the ingredients in place to realize in principle both of the metal-insulator transition scenarios outlined in figure \ref{fig:rgflows}.

From the numerically constructed domain wall solutions, one can read off the asymptotic constants of integration from the behavior of the solutions near the $AdS_5$ boundary as $r \to \infty$. For the case of a massless $B$ field, these take the general form
\bea \label{eq:as_expansion}
& \displaystyle a \; = \; \mu + \frac{\nu}{r^2} + \cdots \,, \quad & \displaystyle w \; = \; \lambda + \frac{\beta  - \lambda \, p^2/2 \log r}{r^2} + \cdots \,, \\
& \displaystyle U  \; = \; r^2 - \frac{\epsilon/3+p^{2}\,\lambda^{2}/6\,\log r}{r^2} + \cdots \,, \quad & \displaystyle v_i \; = \; \log r + \frac{g_i+s_{i}\,\lambda^{2}\,p^{2}/24\,\log r}{r^4} + \cdots \,. \nonumber
\eea
Here $g_{1}+g_{2}+g_{3}=0$ and $s_{1}=s_3=1,\,s_{2}=-2$.
The logarithms indicate the presence of a scaling anomaly. In writing these expansions we have implicitly fixed a scale appearing due to this anomaly. This anomaly is a minor inconvenience, that we treat correctly via holographic renormalization, and will not play a significant role in our discussion.

\marginnote{\sc {\small Phase structure}}[1cm]  The free energy density is then obtained from the renormalized on shell action to be
\be\label{eq:fe}
\frac{\Omega}{V}  = \epsilon - \mu Q - T s \,.
\ee
Where the charge density $Q = - 2 \nu + \k \, p \, \lambda^2/2$. We see that the background `magnetic' field $B$ has shifted the expression for the charge density, due to the Chern-Simons term in the action (\ref{eq:action}). We have also allowed ourselves here to heat up the solutions. The temperature $T$ and entropy density $s$ are defined in the usual way via the surface gravity and the area of the event horizon. An important observable shortly will be the temperature dependence of the conductivity and resistivity. In the supplementary material we describe the standard numerical shooting method that allows us to construct the finite temperature black hole solutions to our model. Including the temperature, our solutions are described by four dimensionful UV parameters: temperature $T$, chemical potential $\mu$, lattice strength $\lambda$ and helical pitch $p$. Taking dimensionless ratios, this gives us a three dimensional phase space to explore.

In the remainder we will focus on the phase structure at low and zero temperature. Firstly we present thermodynamic evidence for a continuous phase transition between zero temperature metallic and insulating phases described by the IR geometries (\ref{eq:ads2}) and (\ref{eq:insulating}) respectively. These phases are not distinguished by any symmetry breaking condensate. The absence of an order parameter is no obstacle in our approach, as we have a precise dynamical characterization of the two phases. In the metallic phase, operators breaking translation invariance are irrelevant at low energies, while in the insulating phase the breaking of translational invariance persists to the lowest energy scales.
This distinction between metals and insulators according to whether ``generalized umklapp'' \cite{Hartnoll:2012rj} processes are irrelevant or relevant at low energies has previously arisen in a soluble one dimensional model \cite{vic}. Below we will compute the optical conductivities and d.c. resistivities of these two phases.

\begin{figure}[h]
\begin{center}
\includegraphics[height = 60mm]{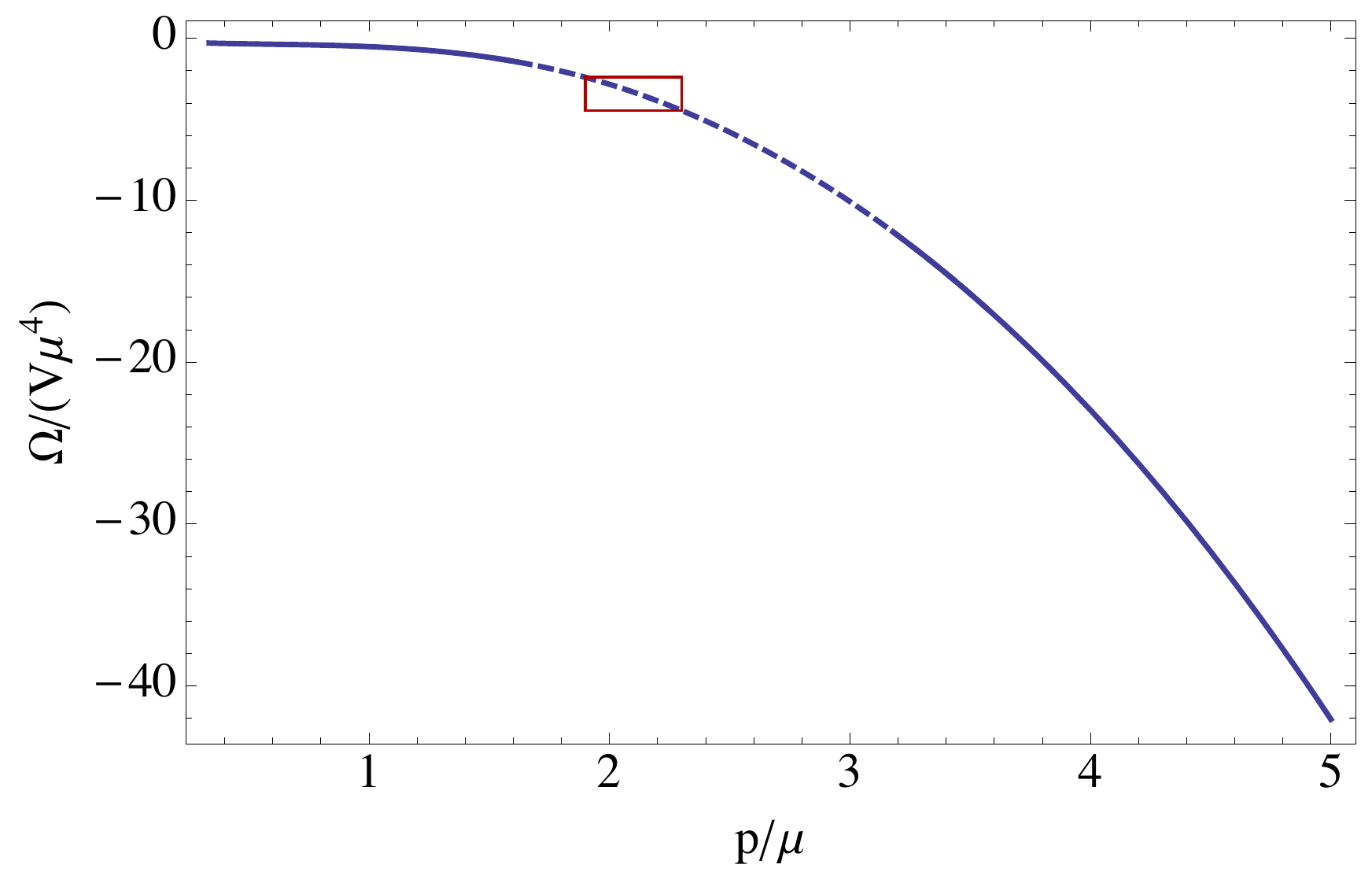}\hspace{0.75cm}\includegraphics[height = 60mm]{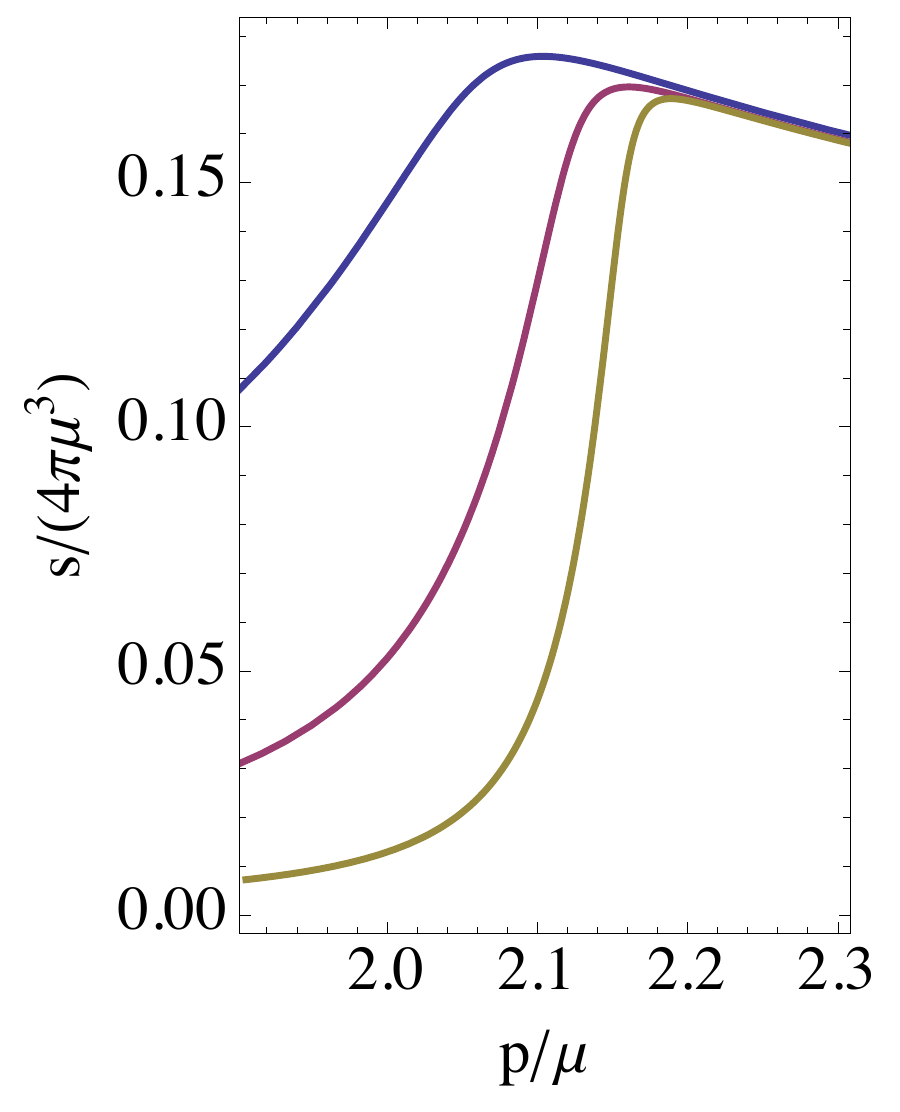}\caption{The left hand plot shows the free energy as a function of pitch. The solid lines are results for $T=0$, while the dashed line shows the thermodynamically preferred phase at $T/\mu = 10^{-3}$. The solid line to the left of the plot corresponds to domain walls with an insulating IR geometry, while the solid line to the right of the plot corresponds to domain walls with metallic IR geometry. There is a continuous quantum phase transition between these two phases. The right hand plot shows the entropy density as a function of the pitch, for values of the pitch close to the critical value $p_c/\mu \approx 2.1$. As the temperature is lowered, temperatures shown from top to bottom are $T/\mu = \{10^{-3},10^{-4},10^{-5}\}$, this curve develops a step function at the critical value. Both plots have $m^2 = 0$, $\k = 1/\sqrt{2}$ and $\lambda/\mu = 3/2$.
\label{fig:phasediag}}
\end{center}
\end{figure}

Figure \ref{fig:phasediag} shows the free energy as a function of the pitch $p$ at a fixed lattice strength and chemical potential. The free energy is computed at a fixed low temperature over the whole range and at zero temperature for some of the range. Issues with numerical accuracy close to the critical point have prevented us from scanning the whole range at zero temperature. The low temperature result is seen to very closely approximate the zero temperature result where the results overlap.
Figure \ref{fig:phasediag} shows that at small pitch the zero temperature limit of the system has the insulating near horizon geometry, while at large pitch the IR geometry is the metallic $AdS_2 \times \R^3$ solution. There is therefore necessarily a quantum phase transition at intermediate helical pitch. We exhibit the phase transition in the figure by plotting the entropy density as a function of the pitch, zoomed in near the putative critical point. The entropy density is a useful quantity to pick out the location of the transition because while the $AdS_2 \times \R^3$ metallic geometry leads to a nonzero entropy density at $T=0$, discussed further below, the insulating geometry (\ref{eq:insulating}) is easily seen to correspond to a low temperature entropy density that vanishes as $s \sim T^{2/3}$.

Figure \ref{fig:phasediag} shows one of many cuts across the phase diagram that contain the physics of interest. We postpone a full exploration of the phase diagram, in particular the $p-\lambda$ plane at zero temperature, to future studies. The fact that we have chosen the parameters in the Lagrangian to be $m^2 = 0$ and $\k = 1/\sqrt{2} \approx 0.7 > 0.57$ implies that the bifurcating solution (\ref{eq:unstable}) does not exist in this instance. Therefore the quantum phase transition will realize the right hand case in figure \ref{fig:rgflows}, and is driven by an operator in the metallic phase becoming relevant. The phase transition appears to be infinite order in this case. It is similar but not identical to the quantum BKT transitions discussed in \cite{Kaplan:2009kr}. The quantum BKT transitions occur as fixed points annihilate at the BF bound. Here instead the dimension of an operator at a fixed point is tuned from irrelevance to relevance (but stays well away from the BF bound). The transition is infinite order because at the critical itself the operator is marginal and therefore the flow away from the unstable point is maximally slow.

\marginnote{\sc {\small Optical conductivity}}[1cm] To actually determine if a given solution describes a metal or an insulator we should of course compute the electrical conductivity.
The frequency dependent conductivity is obtained by perturbing the background by
\be\label{eq:wpert}
\d A = e^{- i \w t}  A(r) \w_1\,, \quad \d B =  e^{- i \w t} B(r) \w_3 \,, \quad \d ds^2 = e^{- i \w t}  \Big(C(r) dt \otimes \w_1 + D(r) \w_2 \otimes \w_3\Big) \,.
\ee
The ordinary differential equations for $\{A,B,C,D\}$ are easy to obtain but somewhat messy. To compute the conductivity we must solve these equations subject to the usual infalling boundary conditions at the horizon \cite{Son:2002sd}. Near the boundary $r \to \infty$, the Maxwell field will behave like
\be
A = A^{(0)} + \frac{A^{(2)} + \half A^{(0)} \w^2 \log r}{r^2} + \cdots \,.
\ee
The logarithm is again indicative of a scaling anomaly in the boundary theory. The (holographically renormalized) electrical conductivity is then, see e.g. \cite{Horowitz:2008bn},
\be\label{eq:sigma}
\sigma = \frac{2}{i\, \w} \frac{A^{(2)}}{A^{(0)}} + \frac{i\, \w}{2} \,.
\ee
This ratio must be taken with all sources for the other modes $\{B,C,D\}$ turned off.
 
Our results for the real -- dissipative -- part of the optical conductivity are shown in figure \ref{fig:opt} below. We show the frequency dependence of the conductivity for several temperatures in both a metallic and an insulating phase. As we anticipated, the metallic phase is characterized by a beautifully conventional Drude peak whose height increases as the temperature is lowered. In the insulating phase, also as anticipated, there is no Drude peak at low temperatures. At intermediately low temperatures we have an incoherent metal, while at the lowest temperatures a soft gap opens as the d.c. conductivity does to zero. Thus we have realized the three phases sketched in figure \ref{fig:cartoon} above. These three behaviors of the conductivity are widely observed in optical studies of metal-insulator transitions \cite{condrev,hussey}.

The power law vanishing of the optical conductivity at zero temperature in the insulating phase may be computed analytically. In the supplementary material we show, using the matching technique of \cite{Donos:2012ra}, that the dissipative conductivity behaves as
\be\label{eq:43}
\text{Re} \, \sigma \; \sim \; \w^{4/3} \,. \qquad (\w \to 0, \quad T = 0, \quad \text{insulating})
\ee
This power law agrees with our numerics. Unlike previously found power law optical conductivities in finite density holography, this power law is not accompanied by a delta function at $\w = 0$. This fact is checked numerically by plotting the imaginary part of the optical conductivity at small frequencies. We are describing an insulating phase. A power law rather than exponential gap in the optical conductivity is presumably related to the persistence of gapless degrees of freedom in the dual theory. These degrees of freedom are manifested in our observation above that the low temperature entropy density scales like a power law, $s \sim T^{2/3}$, in the insulating phase. We discuss this fact further at the end of the paper.

The second notable feature of the insulating plots in figure \ref{fig:opt} is that the spectral weight from the translation-invariance delta function has been transferred both up to `interband' scales of $\w \sim \mu$ -- more specifically, the spectral weight redistribution occurs among frequencies $\w \lesssim 3 \mu$ -- and also into a `mid-infrared peak'. Such major spectral weight transfer, including in some cases to intermediate frequency peaks, is an experimental hallmark of correlation-driven metal-insulator transitions in the cuprates and other materials \cite{condrev, hussey, orenstein, nicoletti}.
The intermediate peak associates an energy scale to the insulating phase. In cases where the metal-instulator transition is continuous, a more comprehensive study of conductivites across the phase diagram should see this energy scale collapse at the critical point. The metallic phase is featureless at intermediate energy scales.

\marginnote{\sc {\small d.c. resistivity}}[1cm] Figure \ref{fig:dc} shows the d.c.{}\,resistivity $\rho = 1/\sigma$ as a function of temperature for two insulating and three metallic systems. At low temperatures the resistivity increases as a function of temperature in the metallic phase and decreases in the insulating phase. Indeed these are the defining features of metals and insulators, respectively!

The low temperature behavior of the plots in figure \ref{fig:dc} can be understood analytically. In the insulating phase we expect the resistivity to diverge as $\rho \sim T^{-4/3}$, corresponding to the scaling found in (\ref{eq:43}). In the metallic phase, the analysis of \cite{Hartnoll:2012rj} implies that the resistivity will go to zero like $\rho \sim T^{2 \Delta(p)}$, where $\Delta(p)$ is the smaller of the irrelevant exponents $\delta_+$ in (\ref{eq:delta}). This is the least irrelevant IR operator responsible for breaking translation invariance.

The UV completion of our boundary theory is a 3+1 dimensional CFT. Therefore, at sufficiently high temperatures dimensional analysis requires that the resistivity decrease like $1/T$ in all phases. This feature can be seen in our numerical results, not shown here, but is quite independent of the universal IR dynamics that is our primary interest. It has the significant consequence, however, that in the metallic phases there is a maximal resistivity at intermediate temperatures. While this point remains to be systematically investigated, there does not seem to be a bound on how large the resistivity can get while the system remains metallic (that is, while the resistivity is still growing with temperature). We therefore obtain, to our knowledge for the first time, a controlled demonstration of the theoretical possibility of bad metals once a quasiparticle picture of transport is abandoned \cite{Emery:1995zz, hussey}. The tight connection we find between bad metals and metal-insulator transitions resonates strongly with the experimental facts \cite{hussey, gun, lim}.

We have demonstrated that interaction-driven metal-insulator transitions can be realized with relative ease in a holographic framework. This fact opens many exciting possibilities for connecting holography with some of the deeper challenges in contemporary condensed matter physics. Within our model, it remains to complete our exploration of the phase diagram and exhibit the other class of bifurcating quantum phase transitions, perhaps as a function of the lattice strength $\lambda$. Looking beyond our model, let us mention two natural, concrete extensions of this work.

\marginnote{\sc {\small Extensions of our model}}[1cm]  Our metallic phase had a zero temperature IR geometry described by $AdS_2 \times \R^3$. It is well known that such phases exhibit a ground state entropy density and much of their interesting `semi locally critical' \cite{Iqbal:2011in} phenomenology may therefore be an artifact of the large $N$ limit of holography. It has recently been emphasized, however, that semi local criticality can also easily arise in holographic models without a ground state entropy density \cite{Gouteraux:2011ce, Hartnoll:2012wm, Anantua:2012nj}. In particular, these models will continue to exhibit power law resistivities as a function of temperature due to lattice scattering in the metallic phase \cite{Hartnoll:2012rj, Anantua:2012nj}. The essential physics we have discussed can be adapted to these models and it is of interest to see how this affects the onset of insulating behavior.

An interesting if exotic feature of our model is that the insulating phases exhibited a soft gap in the optical conductivity, with a power law suppressed spectral weight (\ref{eq:43}). A stronger suppression of spectral weight could be achieved if the full theory were gapped in the insulating phase, for instance by confining the $SU(N)$ gauge fields of the dual field theory. The suggestion that holographic confining phases should describe Mott insulators was made in e.g. \cite{Nishioka:2009zj, Charmousis:2010zz, Balasubramanian:2010uw}. These papers did not break translation invariance and therefore did not exhibit insulating behavior. However, it would be natural to adapt our analysis to a system in which the onset of insulating behavior was tied to a confining phase transition. Indeed it seems plausible that gapped theories with broken translational invariance will always be insulating. Related to this point, it is also of interest to understand to what extent insulating behavior can be decoupled from the appearance of cohesive charge carriers \cite{Hartnoll:2012ux}.

The insulators obtained in this paper do not appear to be Mott-like in the sense that the insulating phases are not tied to charge commensurability. This raises the possibility that we have described a qualitatively new type of strongly correlated gapless insulator.

\section*{Acknowledgements}

It is a pleasure to acknowledge helpful discussions with Jerome Gauntlett, Gary Horowitz, Steve Kivelson, Subir Sachdev and David Tong. S.A.H. is partially supported by a Sloan research fellowship and by a DOE Early Career Award.

\newpage

\section*{Supplementary material}

\subsection*{The equations of motion for the background}

The equations of motion following from the bulk action (\ref{eq:action}) on the metric and vector field ansatz (\ref{eq:metric}) are
\bea\label{eq:eom}
0 & = & a'' + a' \left(v_1' + v_2' + v_3' \right) + p \, \k \, e^{-(v_1 + v_2 + v_3)} w w' \,, \\
0 & = & w'' + w' \left(\frac{U'}{U} + v_1' - v_2' + v_3' \right) - \left(m^2 + p^2 e^{-2(v_1-v_2+v_3)}
- p \, \k \, a' e^{-(v_1 - v_2 + v_3)} \right) \frac{w}{U} \,, \nonumber \\
0 & = & a'^2 - e^{- 2 v_2} U w'^2 + 2 U'(v_1' + v_2' + v_3')
+ 4 U (v_1' v_2' + v_1' v_3' + v_2' v_3')  \nonumber \\
& & -24 + 4 \, p^2 e^{-2 v_1} \sinh^2(v_2 - v_3) + w^2 \left(m^2 e^{-2 v_2} + p^2 e^{-2(v_1 + v_3)} \right)
\,, \nonumber \\
0 & = & 4 ( v_2'' + v_1'') + \frac{2 U''}{U}+e^{-2 v_2} w'^2 - \frac{a'^2}{U}
+ 4 \left(\frac{U'}{U} \left(v_1' + v_2' \right) + v_1'^2 + v_2'^2 + v_1' v_2' \right) \qquad
\nonumber \\
& & - \frac{24}{U} + \left(m^2 e^{-2 v_2} - p^2 e^{-2(v_1 + v_3)} \right) \frac{w^2}{U}
- \left(2 e^{-2 v_1} + e^{-2(v_1 - v_2 + v_3)} - 3 e^{-2(v_1 + v_2 - v_3)}  \right) \frac{p^2}{U}
 \,, \nonumber \\
0 & = & 4 ( v_3'' + v_1'') + \frac{2 U''}{U}-e^{-2 v_2} w'^2 - \frac{a'^2}{U}
+ 4 \left(\frac{U'}{U} \left(v_1' + v_3' \right) + v_1'^2 + v_3'^2 + v_1' v_3' \right) \qquad
\nonumber \\
& & - \frac{24}{U} - \left(m^2 e^{-2 v_2} - p^2 e^{-2(v_1 + v_3)} \right) \frac{w^2}{U}
- \left(2 e^{-2 v_1} - 3 e^{-2(v_1 - v_2 + v_3)} + e^{-2(v_1 + v_2 - v_3)}  \right) \frac{p^2}{U}
 \,, \nonumber \\ 
0 &= & 4 ( v_3'' + v_2'') + \frac{2 U''}{U}+e^{-2 v_2} w'^2 - \frac{a'^2}{U}
+ 4 \left(\frac{U'}{U} \left(v_3' + v_2' \right) + v_3'^2 + v_2'^2 + v_3' v_2' \right) \qquad
\nonumber \\
 &  & - \frac{24}{U} + \left(m^2 e^{-2 v_2} - p^2 e^{-2(v_1 + v_3)} \right) \frac{w^2}{U}
- \left(- 2 e^{-2 v_1} + e^{-2(v_1 - v_2 + v_3)} + e^{-2(v_1 + v_2 - v_3)}  \right) \frac{p^2}{U}
\,. \nonumber
  \eea
 There is a further second order equation of motion that follows from the above equations. Note in particular that the third equation above is a first order equation.
 
 \subsection*{Finite temperature domain walls} 
 
 We demand the existence of a regular black hole horizon in the geometry at the radius $r=r_{+}$ at which $U(r_{+})=0$ and the temperature $T=U'(r_{+})/\left(4\pi\right)$ is finite. We are led to the analytic expansion of our functions
 \begin{align}\label{eq:nh_exp}
 U=&U_{(1)}\,\left(r-r_{+}\right)+\cdots \,, \nn
 w=&w_{(0)}+w_{(1)}\,\left(r-r_{+}\right)+\cdots \,, \nn
 a=&a_{(0)}\,\left(r-r_{+}\right)+\cdots \,, \nn
v_{i}=&v_{i\,(0)}+v_{i\,(1)}\,\left(r-r_{+}\right)+\cdots,\quad i=1,2,3 \,,
 \end{align}
where dots denote higher order terms in $r-r_{+}$. All of the expansion coefficients in \eqref{eq:nh_exp} are determined in terms of the six constants of integration $\{w_{(0)},a_{(0)},v_{i(0)},v_{1(1)}\}$ via the equations of motion. In general we will need 11 constants of integration for a unique solution of the differential system \eqref{eq:eom}. On the other hand, in our asymptotic expansion \eqref{eq:as_expansion} we find another five independent constants of integration, for given values of the sources at the boundary, that can be taken to be $\{\nu,\beta,\epsilon,g_{1},g_{2} \}$. This gives a total of 11 constants of integration for a given $r_{+}$, which we can keep unfixed in order to move through solutions of different temperature.

After fixing $r_{+}$ and the deformation parameters $\{\mu, \lambda, p\}$ we now have a two point boundary value problem to solve. This will give at most a discrete set of solutions for the 11 constants $\{w_{(0)},a_{(0)},v_{i(0)},v_{1(1)},\nu,\beta,\epsilon,g_{1},g_{2}\}$. We have solved for these constants using a shooting method.

 \subsection*{Domain wall solutions for insulating and metallic phases}
 
The zero temperature domain walls are constructed in the same way as the finite temperature solutions just discussed. The only difference is the form of the near-horizon expansions. \\
 
\noindent {\underline{The insulating IR}} \\
\noindent In expanding our solutions in the IR we find it is necessary to use a slightly different coordinate choice to that presented in the main text. Relative to (\ref{eq:insulating}) we let $r \to r + r_+$. The expansion we use for our fields close to the insulating IR solution then reads
 \begin{align}\label{eq:ins_exp}
U=&\,L^{2}(r-r_{+})^{2}\,\left(1+\sum_{j}d^{U}_{j}\,(r-r_{+})^{j/3}+\sum_{j} c_{U}^{j}\, (r-r_{+})^{\delta_{i}}+\cdots\right) \,, \nn
v_{i}=&v_{i(0)}+p_{i}\,\ln (r-r_{+})+\sum_{j}d^{v_{i}}_{j}\,(r-r_{+})^{j/3}+\sum_{j} c_{v_{i}}^{j}\, (r-r_{+})^{\delta_{i}}+\cdots \,, \nn
a=&(r-r_{+})\,\left(\sum_{j}d^{a}_{j}\,(r-r_{+})^{j/3}+\sum_{j} c_{a}^{j}\, (r-r_{+})^{\delta_{i}}+\cdots\right) \,, \nn
w=&w_{(0)}+(r-r_{+})^{2/3}\,\left(\sum_{j}d^{w}_{j}\,(r-r_{+})^{j/3}+\sum_{j} c_{w}^{j}\, (r-r_{+})^{\delta_{i}}+\cdots\right) \,,
 \end{align}
where $L=3\sqrt{2/5}$, $p_{1}=-1/3$, $p_{2}=2/3$, $p_{3}=1/3$, $e^{v_{3(0)}}=2e^{v_{1(0)}+v_{2(0)}}/p$ and $w_{(0)}=4\sqrt{3}\,e^{2 v_{1(0)}+v_{2(0)}}/p^{2}$. In the above expansion the constants $d$ represent the background geometry that governs the near horizon limit and are uniquely fixed by solving the equations of motion\footnote{The expansion \eqref{eq:ins_exp} will be true in general. However, there is a discreet set of values for the CS coupling $\kappa$ for which our expansion \eqref{eq:ins_exp} for the background will be modified by logarithmic terms.}. The $c$'s are constants of integration, of which the three constants $\{c^{1}_{a},c^{2}_{v_{1}},c^{3}_{v_{1}}\}$ can be chosen at will. These are the deformations parameters corresponding to the three irrelevant operators of the IR theory with
\be
\delta_{1}=\frac{1}{6}\,\left(-5+\sqrt{1+120\,\kappa^{2}} \right),\quad\delta_{2}=\frac{1}{6}\,\left(-5+\sqrt{185}\right),\quad\delta_{3}=\frac{1}{6}\,\left(-5+\sqrt{145}\right) \,.
\ee
Finally, the constants $\{v_{1(0)},\,v_{2(0)},r_{+}\}$ that we have included at leading order in the expansion correspond to marginal modes. We therefore have six constants of integration with, as previously, another five constants from our UV expansion (after fixing all deformations and the pitch of the helix). We thus expect to have a unique solution. \\

\noindent {\underline{The metallic IR}} \\
\noindent In the metallic phase we again employ a shift of coordinates $r \to r + r_+$ relative to those used in the main text in (\ref{eq:ads2}). In the case where we have $AdS_{2}\times \mathbb{R}^{3}$ as the IR limit of our zero temperature solutions the near horizon expansion reads
\begin{align}
U=&12\,\left(r-r_{+}\right)^{2}+\sum_{i}c^{U}_{i}\,\left(r-r_{+}\right)^{\delta_{i}+2}+\cdots \,, \nn
v_{i}=&v_{i(0)}+\sum_{i}c^{v_{i}}_{i}\,\left(r-r_{+}\right)^{\delta_{i}}+\cdots \,, \nn
a=&2\sqrt{6}\,\left(r-r_{+}\right)+\sum_{i}c^{a}_{i}\,\left(r-r_{+}\right)^{\delta_{i}+1} \,, \nn
w=&\sum_{i}c^{w}_{i}\,\left(r-r_{+}\right)^{\delta_{i}} \,,
\end{align}
where the constants  $\{c^{U}_{1},c^{w}_{2},c^{v_{2}}_{3}\}$ can be chosen freely and correspond to deformation parameters of the IR theory with
\be
\delta_{1}=1,\quad \delta_{2}=\textstyle -\frac{1}{2}+\sqrt{\frac{1}{4}+\frac{1}{12} p^2 e^{-2v_{1}}-\frac{\kappa}{\sqrt{6}} p e^{-v_{1}}},\quad \delta_{3}=\textstyle -\frac{1}{2}+\sqrt{\textstyle \frac{1}{4} + \frac{1}{3} e^{-2v_{1}}p^{2}} \,.
\ee
These are the exponents quoted in (\ref{eq:delta}) in the main text, here with the mass $m^2 = 0$.
The constants $\{v_{1(0)},v_{2(0)}=v_{3(0)},r_{+}\}$ which appear in the expansion represent marginal modes of the $AdS_{2}\times \mathbb{R}^{3}$ solution. Once again there are six free constants in the IR, leading to a unique solution
 for given deformation parameters in the UV.
 
 \subsection*{Holographic renormalization of action}
 
As usual, our bulk action \eqref{eq:action} needs to be supplemented by appropriate boundary counterterms. In particular we need to add
\begin{align}
S_\text{bdy}=\int d^{4}x\,\sqrt{-\gamma}\,\left[2\,K-6+\frac{1}{4}\,\ln r\,\left(F_{mn}F^{mn}+G_{mn}G^{mn}\right)+\cdots\right].
\end{align}
Here $K=\g^{mn}\nabla_{m}n_{n}$ is the trace of the extrinsic curvature of the boundary, $n^{m}$ is an outward pointing unit normal vector and $\gamma$ is the determinant of the induced metric at infinity. The logarithmic terms cancel the divergence related to the trace anomaly in four dimensions. The ellipsis indicate a Ricci scalar term of the four dimensional boundary metric, which will not contribute given our asymptotics \eqref{eq:as_expansion}.

 \subsection*{Analytic results for the optical conductivity} 

At the lowest frequencies, the dissipative part of the conductivity (\ref{eq:sigma}) can be computed purely from near-horizon IR data. The general matching formula derived in \cite{Donos:2012ra} says that as $\w \to 0$
\be\label{eq:match}
\text{Re} \, \sigma(\w) = \sum_I d^I \frac{\text{Im}\, {\mathcal G}^R_I(\w)}{\w} \,.
\ee
Here the sum is over all operators $\ocal_I$ in the IR theory described by the near horizon geometry. The operators are taken to diagonalize the IR scaling operator. That is, they are decoupled at a quadratic level in the bulk. The $d^I$ are real coefficients that are nonzero if the corresponding operator has an overlap with the current operator $\vec J$ in the UV. The ${\mathcal G}^R_I(\w)$ are the retarded Green's functions of these operators in the IR theory. We must therefore proceed to obtain the Green's functions for the IR insulating geometry (\ref{eq:insulating}).

The equations of motion for the frequency dependent perturbation (\ref{eq:wpert}) about the insulating IR background (\ref{eq:insulating}) are fairly simple. Keeping the lowest powers in $r \to 0$ one finds immediately a decoupled equation for the Maxwell perturbation $A(r)$
\be
A'' + \frac{10}{3 r} A' + \frac{25 \, \w^2}{324 \, r^4} A = 0 \,.
\ee
A further mode is decoupled by defining
\be
E = r \frac{d}{dr} \frac{C}{r^{2/3}} \,,
\ee
which is then found to satisfy
\be
E'' - \frac{5(288 \, r^2 - 5 \, \w^2)}{324 \, r^4} E = 0 \,.
\ee 
These two equations are easily solved in terms of Hankel functions. Imposing infalling boundary conditions at the horizon,
\be
E = \sqrt{\frac{r}{\w}} H^{(1)}_{13/6}\left( \frac{5\w}{18 r} \right) \,, \qquad A = \left(\frac{\w}{r}\right)^{7/6} H^{(1)}_{7/6}\left( \frac{5\w}{18 r} \right) \,.
\ee
Expanding for $r \to \infty$ and taking the ratio of the normalizable by the non-normalizable mode, as required in the matching procedure \cite{Donos:2012ra}, one finds
\be\label{eq:aa}
\text{Im}\,  {\mathcal G}^R_A(\w) \sim \w^{7/3} \,, \qquad \text{Im}\,  {\mathcal G}^R_E(\w) \sim \w^{13/3} \,.
\ee
There is a third physical mode in the IR that is more difficult to decouple. However,
because the IR equations of motion only depend on the ratio $\w/r$, it is possible to extract the power of $\w$ by solving the equations of motion with $\w =0$. This is equivalent to finding the IR dimension of the operator. With $\w = 0$ the remaining IR equation is
\be
B'' + \frac{2}{r} B' - \frac{40}{9 r^2} B = 0 \,.
\ee
This has solution $B = c_1 r^{5/3} + c_2 r^{-8/3}$. Dimensional analysis then implies that this operator will also lead to
\be\label{eq:bb}
\text{Im}\,  {\mathcal G}^R_B(\w) \sim \w^{13/3} \,.
\ee
From (\ref{eq:aa}) and (\ref{eq:bb}) we see that the most singular of the three IR operators is $A$. Thus we conclude from (\ref{eq:match}) that the optical conductivity at small frequencies will scale as
\be
\text{Re} \, \sigma(\w) \sim \w^{4/3} \,.
\ee
This is the result quoted in the main text in (\ref{eq:43}) and observed in our numerical results.
 
\end{document}